# Density functional theory calculations on magnetic properties of actinide compounds

**Denis Gryaznov,**[*a] **Eugene Heifets**[a,d] **and David Sedmidubsky**[b]

We have performed a detailed analysis of the magnetic (collinear and noncollinear) order and atomic and the electron structures of $UO_2$, $PuO_2$ and UN on the basis of density functional theory with the Hubbard electron correlation correction (DFT+$U$). We have shown that the 3-k magnetic
10 structure of $UO_2$ is the lowest in energy for the Hubbard parameter value of $U$=4.6 eV (and $J$=0.5 eV) consistent with experiments when Dudarev's formalism is used. In contrast to $UO_2$, UN and $PuO_2$ show no trend for a distortion towards rhombohedral structure and, thus, no complex 3-k magnetic structure is to be anticipated in these materials.

## 1. Introduction

15 Actinide compounds continue to attract a great interest for both materials scientists and nuclear engineers. Their properties combine a strong electron correlation and relativistic effects of 5f valence electrons. In this paper, we study collinear and non-collinear magnetic structures of three basic actinide materials
20 $UO_2$, $PuO_2$ and UN. All these materials have face-centred cubic (f.c.c.) actinide sub-lattice: the two oxides have fluorite structure and UN rock-salt structure. Experiments suggest that at low temperatures UN is anti-ferromagnetic[1] with a collinear magnetic order, where U magnetic moments alternate along the <001>
25 direction, while $UO_2$ is anti-ferromagnetic (AFM) with the so-called noncollinear 3-k ordering of U magnetic moments (see section 3 for more detail description of different magnetic structures). The U magnetic moments in UN and $UO_2$ in the AFM phases are very different, being 0.75 $\mu_B$ and 1.74 $\mu_B$, respectively.
30 The Néel temperatures for both $UO_2$ ($T_N$ = 30.8 K, ref. 2) and UN ($T_N$ = 53 K, ref. 1) are quite low. These two materials also differ in the chemical bonding. UN is evidently a conductor[3], whereas $UO_2$ is a Mott insulator (as discussed, for example, in Ref. 4). $PuO_2$ is also a Mott insulator[4] with magnetic susceptibility being
35 temperature independent[5]. All recent theoretical considerations[6-9] employing the DFT+$U$ technique or hybrid exchange-correlation functional (though without including spin-orbit interactions (SOI)) suggested the 1-k (collinear) AFM order for insulating $PuO_2$, while experiment suggests that $PuO_2$ is diamagnetic.
40 Thus, it is important to compare magnetic orders and accompanying lattice distortions for three considered compounds ($UO_2$, UN, and $PuO_2$) using the same method. Ignoring the lattice distortions may lead to a wrong electronic structure and significant errors in the defect energetics[10]. As it was already
45 mentioned, these materials reveal the same f.c.c. structure in the actinide sublattice. Therefore, similar structure of exchange interactions could be expected.
  $UO_2$ has been studied most intensively and now is much better understood in a comparison with $PuO_2$ and UN. $UO_2$ is
50 experimentally known to have a transverse 3-k magnetic structure and oxygen sub-lattice distortion of the same symmetry[11]. To the best of our knowledge, the only first-principles modelling of the non-collinear magnetic ordering in $UO_2$ was published by Laskowski et. al[12]. This study employed the DFT+$U$ technique
55 within the local spin density approximation (LSDA)[13] and all-electron linearized augmented plane wave plus local orbitals method (L/APW+lo)[14] as implemented in the Wien-2k computer code. In these computations, the energetic preference of the 3-k structure with respect to a regular 1-k structure was primarily
60 dependent on the method used to correct for a double counting of on-site interactions. The 3-k structure appears to be more stable, if the double counting correction accurately includes spin-polarization of the electron density[11], like it is done in LSDA+U [15,16] or in simplified rotationally-invariant approach by Dudarev
65 et. al.[17]. Nevertheless, the 1-k and 2-k magnetic structures[2,18] were also suggested for $UO_2$ prior to Ref. 11. Also, no significant lattice distortions were found in these early experimental studies of $UO_2$. Only recently, it was shown computationally[19] for collinear AFM ordering in $UO_2$ that the U magnetic moments
70 alternate along the <111> direction, but not along the <001> direction, as it was generally assumed in nearly all previous computer simulations. For the simplicity, we call these structures hereafter as the "<111> magnetic structure" and the "<001> magnetic structure". The study[19] based on the electronic structure
75 calculations with hybrid exchange-correlation functional found that the rhombohedral unit cell has a lower energy than the tetragonal one, even though the SOI are not included. Thus, change from usual <001> magnetic structure to the <111> one could indicate possible non-collinear magnetism.
80 To the best of our knowledge, no such studies on the magnetic properties of $PuO_2$ and UN have been performed so far. X-ray diffraction measurements on UN revealed no significant tetragonal distortion[20], which would be a consequence of the AFM spin alignment along the <001> direction.
85 In the present study, we consider possible collinear and non-collinear magnetic structures of $UO_2$ also using the DFT+$U$ technique, but implemented in another code, Vienna Ab-intio Simulation Package (VASP)[21-22]. First, we test ability of this method and the code to reproduce experimentally observed non-
90 collinear magnetic order in $UO_2$ using experimental values of the Hubbard parameter ($U$ = 4.6 eV, $J$ = 0.5 eV) [24]. Second, we explore different possible magnetic structures in UN and PuO$_2$ using the same DFT+$U$ technique and try to determine which of the <001> and <111> structures is more stable. Section 2 describes computational details used in the present simulations. Descriptions of studied magnetic structures are given in section 3. The results of our computations are provided and discussed in section 4. Lastly, the conclusions are summarized in section 5.

## 2. Computational details

In present first-principles simulations we used the VASP (version 4.6)[21-22] computer code employing the DFT+$U$ method. The VASP code treats core electrons using pseudopotentials, whereas the semi-core electrons at U atoms and all the valence electrons are represented by plane waves. The electronic structure is calculated within the projector augmented wave (PAW) method[23]. The simplified rotationally-invariant Dudarev's form[17] for the Hubbard correction was used for UO$_2$ and UN. It uses exclusively the difference $U_{eff} = U - J$ of the Hubbard parameter $U$ and the exchange parameter $J$. In contrast to uranium compounds, PuO$_2$ shows a significant role of exchange part requiring use of Liechtenstein's form[16] for the energy correction. The double counting correction in all our calculations was treated with account for spin-polarization.[15-17] Computations of UO$_2$ were done including the SOI effects, whereas computations of PuO$_2$ and UN employed only scalar relativistic approximation. Both unit cell parameters and atomic positions were optimized until the energy convergence reached $10^{-5}$ eV. The calculations were performed with the cut-off energy of 520 eV. The integrations in the reciprocal space over the Brillouin zone (BZ) of the tetragonal unit cell of PuO$_2$ and UN (used to calculate the <001> AFM magnetic structure) were performed using 10x10x8 and 12x12x10 Monkhorst-Pack meshes[25], respectively. Computations of the rhombohedral PuO$_2$ and UN with the <111> magnetic structure were performed with 12x12x12 and 14x14x14 Monkhorst-Pack meshes. Similarly, the integrations over the BZ for the conventional unit cell of UO$_2$ were performed using 6x6x6 Monkhorst-Pack meshes. The conventional 12-atom unit cell was necessary for modelling of UO$_2$ with non-collinear magnetic structures. It was possible to use the smaller unit cell for a collinear magnetic ordering (the 1-k AFM <001> and <111> magnetic structures) in UO$_2$. Correspondingly, in these cases we applied larger 14x14x10 and 12x12x12 k-meshes. The applied meshes in the reciprocal space were sufficient to reach convergence of $10^{-4}$ eV for one-electron energies. Fractional electron occupancies were estimated with the Gaussian method using the smearing parameter of 0.25 eV. Calculations, which included SOI, were done with lifted symmetry constraints.

Photoemission spectroscopy (PS) measurements by Baer and Schoenes[24] suggest that the Hubbard correlation parameter $U$ is 4.6 eV for UO$_2$ assuming that exchange parameter $J$ is 0.5 eV. These values were applied later by Dudarev[17]. In their calculations[17] the band gap becomes open and equal to 1.3 eV within the LSDA+$U$, being, however, smaller than the experimental value of 2.0 eV. A somewhat better agreement is observed within the generalized gradient approximation[26], i.e. GGA+$U$[10,27-29]. Note that following Dudarev's calculations, we employed recently the same values of $U$ and $J$ in our study on bulk properties and defects behaviour in UO$_2$[10]. In the present simulations we used the same set of correlation $U$ and exchange $J$ parameters for computations of UO$_2$. The parameter $U_{eff}$ =1.875 eV for UN was fitted[30] to reproduce the magnetic moment of uranium ions and UN unit cell volume in the low-temperature phase. The band gap of ~1.8 eV[31] for PuO$_2$ is known from the electrical conductivity measurements what is similar to the band gap in UO$_2$. Previous theoretical studies[6-9] also agreed on the AFM solution for PuO$_2$ within the 1-k magnetism and, therefore, used the tetragonal structure as described above. Despite the relatively similar band gaps in both oxides, their electronic structures are quite different what is clearly seen in the corresponding PS measurements[32]. Parameters $U$=3.0 eV and $J$=1.5 eV were fitted for PuO$_2$ to describe correctly its experimental lattice constant, band gap, position of Pu 5f band and the magnetic moment on Pu atoms.

## 3. Magnetic structures

The dependence of atomic magnetic moments on the position in a lattice can be expressed as expansion in plane waves:

$$\mathbf{M}_j = \sum_{w=1}^{k} e^{i\mathbf{k}_w(\mathbf{r}_j - \mathbf{r}_0)} \mathbf{M}_0^w,$$

where $\mathbf{M}_j$ is magnetic moment of the atom in unit cell j and at position $\mathbf{r}_j$, $\mathbf{r}_0$ is the position of the same atom in the 0$^{th}$ unit cell, $\mathbf{k}_w$ and $\mathbf{M}_0^w$ are, respectively, the wave vector and amplitude of the magnetic wave $w$.

In the collinear 1-k magnetic structures magnetic moments of U atoms are collinear and changes in the magnetic moments can be described by a single wave ($k$=1). For the <001> magnetic structure choosing the Oz axis along the direction of alternation of magnetic moments, the wave vector is $\mathbf{k}_1$= 2π/$a$ (0, 0, 1), where $a$ is a cubic lattice constant. Similarly, for the <111> structure the wave vector is $\mathbf{k}_1$= π/a (1, 1, 1). These two collinear 1-k magnetic structures were modelled for all three materials considered here. These magnetic structures have symmetry reduced from the cubic one. In the <001> structure the lattice has a tetragonal symmetry, and in the <111> structure the lattice becomes rhombohedral, as can be seen from the next section.

Farber and Lander[18] suggested the 2-k transverse magnetic structure for UO$_2$ which associated with a transverse phonon. If we choose the direction of the phonon propagation as the Oy axis, then magnetic waves propagate along the Ox and Oz axes ($\mathbf{k}_1$= 2π/a (1, 0, 0), $\mathbf{k}_2$= 2π/a (0, 0, 1) ) with amplitudes $\mathbf{M}_0^1$= M$_0$ (0, 1, 0), $\mathbf{M}_0^2$= M$_0$ (1, 0, 0), where M$_0$ is magnitude of atomic magnetic moment. Magnetic moments of U atoms lie on the Oxy plane and point along various [110] directions. The transverse phonon in this structure can be described as O atoms in odd and even {010} oxygen planes shift in opposite directions along the Ox axis. While later experiments showed that this structure is not the most stable one, we included it into our simulations to compare energies of all previously considered magnetic structures of UO$_2$.

According to the experiment[11], UO$_2$ has transverse 3-k magnetic structure. The wave vectors for three waves in 3-k structures are $\mathbf{k}_1$= 2π/a (1, 0, 0), $\mathbf{k}_2$= 2π/a (0, 1, 0), $\mathbf{k}_3$= 2π/a (0, 0, 1). There are two equivalent transverse structures with this symmetry in the fluorite lattice. The first structure has amplitudes $\mathbf{M_0}^1 = M_0 (0, 1, 0)$, $\mathbf{M_0}^2 = M_0 (0, 0, 1)$, $\mathbf{M_0}^3 = M_0 (1, 0, 0)$. The second one has amplitudes $\mathbf{M_0}^1 = M_0 (0, 0, 1)$, $\mathbf{M_0}^2 = M_0 (1, 0, 0)$, $\mathbf{M_0}^3 = M_0 (0, 1, 0)$. The two O atoms nearest to each U atom in the direction of its magnetic moment shift from their sites toward this U atom. Both structures have the same total energies. We used the first one in our simulations.

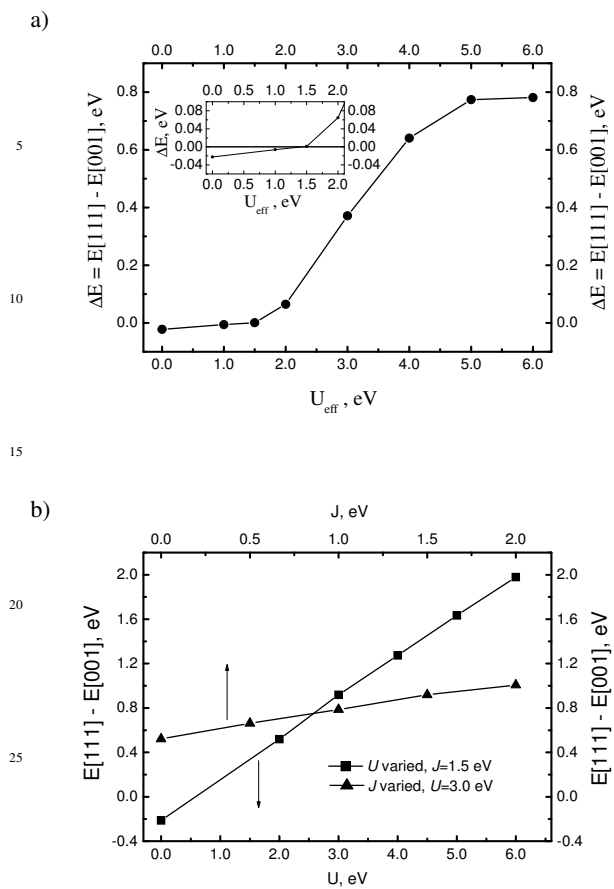

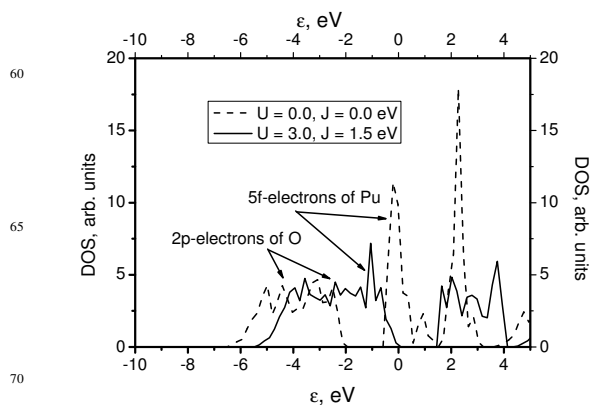

Fig. 2 A comparison of total density states (DOS) for $PuO_2$ for $U$ = 0.0, $J$ = 0.0 eV (dashed line) and $U$ = 3.0, $J$ = 1.5 eV (solid line). The DOS was calculated by employing the tetrahedron method[29] with given occupations of the electron states. The Fermi energy is taken as zero; ε is one-electron energy.

Fig. 1 The energy difference between the <111> and <001> magnetic structures for (a) UN as a function of $U_{eff}$ (Dudarev's functional); the inset contains enlarged fragment of the same plot at $U_{eff} \leq 2.0$ eV; (b) $PuO_2$ as functions of one of the $U$ and $J$ parameters (Lichtenstein's functional), while another parameter is fixed.

## 4. Results and discussion

In the present study, we assess the difference between the two <111> and <001> magnetic collinear structures, as a function of the $U$ and $J$ parameters for UN and $PuO_2$ (fig. 1).

The energy difference between the two magnetic structures for UN (fig. 1a) is very small and negative at small values of $U_{eff} = U-J$. It slowly grows for $U_{eff}$ between 0.0 eV and 1.5 eV, then noticeably increases from 2.0 to 5 eV, and likely saturates for the higher values of $U_{eff}$. For the optimized value of $U_{eff} = 1.875$ eV the <001> structure of UN is already more stable than the <111> structure (see inset in fig. 1b). At this value of $U_{eff}$ the lattice constants for UN in the <001> structure are $a$=4.974 Å and $c$=4.859 Å, and lattice parameters in the <111> structure are $a$=4.942 Å and γ=88.2°. In both cases the cubic unit cell is distorted along the direction of alternation of magnetic moments. In the <001> structure it is compressed along the Oz axis, for the <111> structure the unit cell is elongated along [111] direction. It is experimentally known that UN is cubic with the lattice constant $a$=4.886 Å[33]. The calculated spin moments on U atoms are 1.47 $\mu_B$ in the <001> structure and 1.82 $\mu_B$ in the <111> structure. The magnetic moment of U atoms measured[1] at low temperatures is 0.75$\mu_B$. Inclusion of SOI in calculations allows revealing substantial orbital moments in actinide compounds what would lead to much better alignment of U atom magnetic moment with experimental value[30].

Due to the Liechtenstein form of the DFT+U functional[16] applied to $PuO_2$, we have to vary the $U$ and $J$ parameters independently. It was done by varying $U$ with the $J$-parameter fixed at 1.5 eV and by varying $J$ at $U$= 3.0 eV, correspondingly. As seen in fig. 1b, the <001> magnetic structure of $PuO_2$ is energetically more stable than the <111> one, except for very small values of Hubbard parameter $U$. It suggests no preference of the <111> magnetic structure, in contrast to $UO_2$ (see discussion below), for realistic values of $U$ and $J$ parameters. The difference increases with both parameters, indicating further stabilization of the <001> magnetic structure in a comparison to the <111> one. The energy difference between the two magnetic structures (fig. 1b) is almost linear for $PuO_2$, independently of which parameter is varied or fixed. For chosen values of the parameters ($U$ = 3.0 eV and $J$ = 1.5 eV), lattice constants for $PuO_2$ in the <001> structure are $a$=5.402 Å and $c$=5.513 Å, and lattice parameters in <111> structure are $a$=5.430 Å and γ=88.9°. In the case of $PuO_2$ a cubic unit cell becomes elongated in the direction of alternation of magnetic moments. The calculated spin moments at Pu atoms are 3.81$\mu_B$. Experimentally, $PuO_2$ is cubic with lattice constant $a$=5.398 Å[34] and diamagnetic[5].

In fig. 2 we present the total densities of states (DOS) for the discussed tetragonal AFM unit cell of $PuO_2$, when the strong correlation effects are neglected (dashed line) and for the employed values of $U = 3.0$ eV and $J = 1.5$ eV (solid line). The DOS clearly demonstrates that PuO$_2$, like UO$_2$, tends to be metallic if the strong correlation effects are not treated properly, whereas the band gap of 1.5 eV appears for the chosen parameters of the GGA+$U$ scheme. The latter value of the band gap is slightly smaller than the experimental value (1.8 eV).

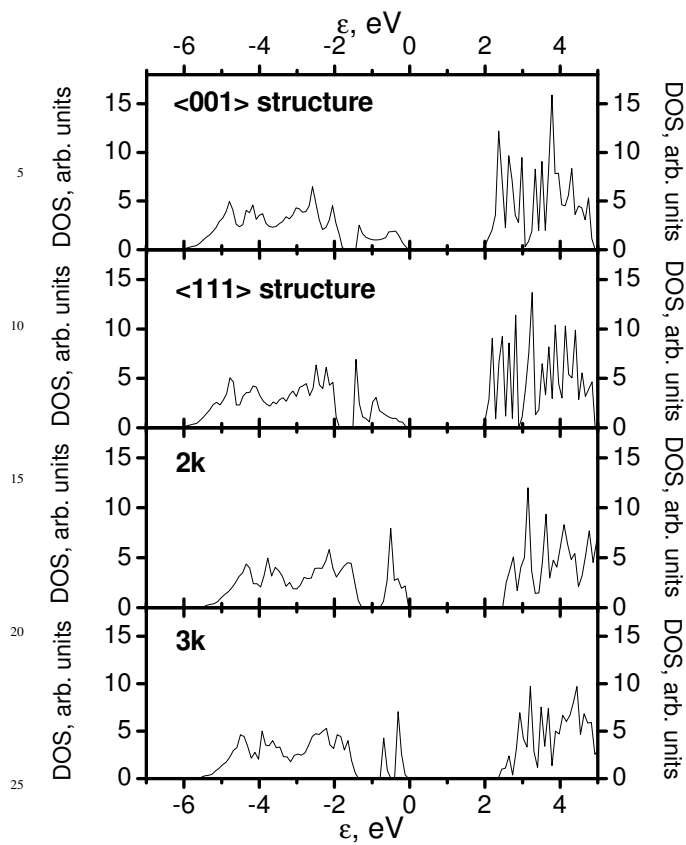

Fig. 3. The total DOS of UO$_2$ calculated for <001>, <111>, 2k and 3k structures. The Fermi energy is taken as zero; ε is one-electron energy.

The case of UO$_2$ differs from the discussed above trends for UN and PuO$_2$, reflecting the fact that the <111> magnetic structure in UO$_2$ is more stable than the <001> one by 62 meV/formula unit at $U = 4.6$ eV and $J = 0.5$ eV. This result confirms previously published the hybrid functional calculations[19] with atomic basis set. Due to the SOI the total energy is reduced almost by 2.66 eV per UO$_2$ primitive unit cell. This does not affect relative energies of all studied magnetic structures (3k, 2k and both <001> and <111> 1k structures). Relative energies for all considered magnetic structures are provided in Table 1 with respect to the 3k magnetic structure. The transverse 3-k magnetic structure appears to be the most energetically preferable. This is in accord with inelastic neutron scattering experiments[11]. The 2-k structure proposed by Faber and Lander[18] has just a little bit lower energy (5 meV/formula unit) than the <111> collinear structure but noticeably higher than the transverse 3-k structure.

Both the <001> and <111> collinear structures have unit cells compressed along the direction of alternation of magnetic moment (see Table 1). Magnetic moments of U atoms in both structures point in the same [001] and [111] directions.

All lattice constants in the 2-k structure are different. The lattice of the 2-k structure becomes orthorhombic. As expected (see Section 3 and Ref. 18), odd and even oxygen {010} planes are shifted along the Ox axis in the opposite directions. The obtained shift is $\Delta = 9.7 \cdot 10^{-3} a$ (compare with $\Delta = 2.6 \cdot 10^{-3} a$ obtained in Ref. 10). However, directions of magnetic moments are very different from those suggested in Ref. 11: the magnetic moments point almost along the [010] directions, but are slightly tilted towards shorter square diagonal (the squares are perpendicular to the [001] direction). This can be expressed by amplitudes of magnetic waves $\mathbf{M_0}^1 = (0, 1, 0)*1.79$ μ$_B$, $\mathbf{M_0}^2 = (1, 0, 0)*0.24$ μ$_B$.

Unit cell in the transverse 3-k structure keeps cubic shape. Magnetic moments are aligned according the transverse 3-k symmetry. The pair of O atoms nearest to each U atom in the direction of its magnetic moment is shifted toward this U atom by $9.6*10^{-3}\sqrt{3}\, a$ (or 0.092 Å).

For the total magnetic moments on U atoms in the most stable transverse 3-k structure we obtained value of 1.99 μ$_B$, which slightly exceeds the experimental value of 1.74 μ$_B$. Magnetic moments obtained for the <001> 1-k and for the 2-k structures are much closer to the experimental value, but these structures have higher energies and are not consistent with inelastic neutron scattering data[11].

In fig. 3 we compare the total DOS for different magnetic UO$_2$ structures. In all considered structures, the highest valence band consists predominantly of U 5f orbitals and the next highest valence band is mostly built from O 2p orbitals. The conduction bands contain U 6d and U 5f orbitals. Our calculations re-produce the band gaps in various magnetic structures of UO$_2$ (see Table 1) very close to the experimental value (2.0 eV)[24]. The band gaps in both considered non-collinear structures are a little larger, by several tenths of eV. In calculations[19] with hybrid functional the band gap for the <111> structure is significantly (by ~1.5 eV) overestimated.

Table 1. Results of calculations for UO$_2$. ΔE is the total energy (in meV/molecule) for various magnetic structures in UO$_2$ with respect to transverse 3-k structure, which has the lowest energy. The energy calculations included SOI. $a,b,c$ are lattice constants, α, β and γ are angle between lattice vectors of conventional unit cell. E$_g$ is band gap. μ is the total magnetic moment of U atom (in Bohr's magnetons μ$_B$) and the values in parentheses are spin contributions to the magnetic moments. The experimental value of magnetic moment is 1.74 μ$_B$.

|  | Magnetic structure | | | |
|---|---|---|---|---|
|  | <001> 1-k | <111> 1-k | Faber-Lander 2-k | Transverse 3-k |
| ΔE, meV/molecule | 95 | 33 | 28 | 0 |
| a, Å | 5.566 | 5.550 | 5.555 | 5.547 |
| b, Å | 5.566 | 5.550 | 5.562 | 5.547 |
| c, Å | 5.508 | 5.550 | 5.521 | 5.547 |
| α=β=γ, ° | 90 | 91.7 | 90 | 90 |
| E$_g$, eV | 1.95 | 2.03 | 2.50 | 2.38 |
| μ, μ$_B$ | 1.76 (1.95) | 2.00 (1.98) | 1.81 (2.04) | 1.99 (2.00) |

The U 5f band width is 1.5 eV for the collinear magnetic structures and gets much narrower for the non-collinear cases (0.76 eV and 0.86 eV in case of 2-k and 3-k structures, respectively). This band splits into two separate subbands: U 5f(5/2) and U 5f(7/2) in the 3-k structure with a gap of ~0.1 eV and distance between peaks ~0.38 eV. The width (~4.2 eV for <111> structure and ~4.4 eV for other structures) of O 2p band varies little among considered structures. The gap between O 2p and U 5f valence bands is small, ~0.3-0.5 eV . As a result, O 2p band shifts, following the narrowing of U5f valence band, and becomes by ~0.5 eV closer to the Fermi level in the non-collinear structures than in the collinear ones.

## Conclusions

We have compared several possible magnetic structures of several key actinides $UO_2$, UN and $PuO_2$ based on the GGA+$U$ technique. Our modelling shows that the transverse non-collinear 3-k structure of $UO_2$ is the most stable one for this material. $UO_2$ retains a cubic shape in this structure. Two O atoms nearest to each U atom in the direction of its magnetic moment move toward this U atom. This is consistent with both experiment[11] and previous computer simulation[12] employing the LDA+$U$ technique within the Wien2k code. It is important that such agreement is achieved with the standard values of Hubbard and exchange parameters ($U$=4.6 eV, $J$=0.5 eV) within Dudarev's form of the DFT+$U$ approach[17]. Still, a reason for overestimated U atom magnetic moment remains unclear.

The collinear magnetic order causes breaking of cubic symmetry in UN and $PuO_2$ . In contrast to $UO_2$, neither UN nor $PuO_2$ show the energetical preference for the rhombohedral distortion. Both materials have the AFM tetragonal <001> structure for a reasonable choice of parameters $U$ and $J$. The total DOS of $PuO_2$ is successfully reproduced using the Liechtenstein form[16] for the Hubbard correction with the parameters $U$ = 3.0 eV and $J$ = 1.5 eV. However, as well as in the previous computational studies[6-9], we obtained that AFM state of $PuO_2$ is more stable than the experimentally observed diamagnetic state[5].
.

## Acknowledgements

This study was supported by the F-BRIDGE project, as part of the 7[th] EC Framework Programme, and the Proposal Nr. 25592 from the EMS Laboratory of the PNNL. Authors are greatly indebted to E. Kotomin, R. Caciuffo, R. A. Evarestov, R. Konings for very useful discussions. We acknowledge also the EC for support in the frame of the Program "Training and mobility of researchers".

## Notes and references

[a] *European Commission, Joint Research Centre, Institute for Transuranim Elements, Postfach 2340, Karlsruhe, D-76125, Germany.*

[b] *Institute for Solid State Physics, University of Latvia, Kengaraga 8, LV-1063,Riga, Latvia.* E-mail: *gryaznov@mail.com*

[c] *Institute of Chemical Technology, Technicka 5, 16628 Prague, Czech Republic,* E-mail: *David.Sedmidubsky@vscht.cz*

[d] E-mail: *EHeif5719@sbcglobal.net*

1. N. A. Curry, *Proc. Phys. Soc.,* 1965, **86**, 1193
2. B. C. Frazer, G. Shirane, D. E. Cox, *Phys. Rev.,* 1965, **140 (4A)**, A 1448
3. M. Marutzky, U. Barkow, J. Schoenes, R. Troć, *J. Mag. Mag. Mat.*, 2006, **299**, 225
4. I. D. Prodan, G. E. Scuseria, R. L. Martin, *Phys. Rev. B,* 2007, **76**, 033101
5. M. Colarieti-Tosti, O. Eriksson, L. Nordström, J. Wills, M. S. S. Brooks, *Phys. Rev. B,* 2002, **65**, 195102
6. I. D. Prodan, G. E. Scuseria, J. A. Sordo, K. N. Kudin, R. L. Martin, *J. Chem. Phys.,* 2005, **123**, 014703
7. G. Jomard, B. Amadon, F. Bottin, M .Torrent, *Phys. Rev. B,* 2008, **78**, 075125
8. B. Sun, P. Zhang, X.-G. Zhao, *J. Chem. Phys.,* 2008, **128**, 084705
9. F. Jollet, G. Jomard, B. Amadon, J. P. Crocombette, D. Torumba, *Phys. Rev. B,* 2009, **80**, 235109
10. D. Gryaznov, E. Heifets, E. Kotomin, *Phys. Chem. Chem. Phys.,* 2009, **11,** 7241
11. R. Caciuffo, G. Amoretti, P. Santini, G. H. Lander, J. Kulda, P. de V. Du Plessis, *Phys. Rev. B,* 1999, **59**, 13892
12. R. Laskowski, G. K. H. Madsen, P. Blaha, K. Schwarz, *Phys. Rev. B,* 2004, **69**, 140408
13. W. Kohn, L .J. Sham, *Phys. Rev.,* 1965, **140,** A1133
14. P. Blaha, K. Schwarz, G. K.H. Madsen, D. Kvasnicka, J. Luitz, WIEN2k, *An Augmented Plane Wave + Local Orbitals Program for Calculating Crystal Properties* (K. Schwarz, Techn. Universität Wien, Austria), ISBN 3-9501031-1-2, 2001
15. M. T. Szyżyk, G. A. Sawatzky, *Phys. Rev. B,* 1994, **49 (20),** 14211. The technique, proposed in this paper, is often called "around mean field" method. Under this name it is used in Wien2k code[14] and in ref. 12.
16. A. I. Liechtenstein, V. I. Anisimov, J. Zaane, *Phys. Rev. B,* 1995, **52**, R5467
17. S. L. Dudarev, G. A. Botton, S. Y. Savrasov, C. J. Humphreys, A. P. Sutton, *Phys. Rev. B,* 1998, **57 (3)**, 1505
18. J. Faber, Jr, G. H. Lander, *Phys. Rev. B,* 1976, **14 (3)**, 1151
19. R.A. Evarestov, A. Bandura, E. Blokhin, *Acta Mater.,* 2008, **57**, 600
20. H. W. Knott, G. H. Lander, M. H. Mueller, O. Vogt, *Phys. Rev. B,* 1980, **21 (9)**, 4159
21. G. Kresse, J. Furthmüller, *Comp. Mater. Sci.,* 1996, **6**, 15
22. G. Kresse, J. Furthmüller, *VASP, the Guide* (University of Vienna), 2007
23. G. Kresse, D. Joubert, *Phys. Rev. B,* 1999, **59 (3)**, 1758
24. Y. Baer, J. Schoenes, *Solid State Commun.,* 1980, **33**, 885
25. J. Monkhorst, J. D. Pack, *Phys. Rev. B,* 1976, **13**, 5188
26. J. P. Perdew, K. Burke, M. Ernzerhof, *Phys. Rev. Lett.,* 1996, **77**, 3865
27. H. Y. Geng, Y. Chen, Y. Kaneda, M. Kinoshita, *Phys. Rev. B,* 2007, **75**, 054111
28. M. Iwasawa, Y. Chen, Y. Kaneta, T. Ohnuma, H.-Y. Geng, M. Kinoshita, *Mater. Trans.,* 2006, **47 (11)**, 2651
29. F. Gupta, G. Brillant, A. Pasturel, *Phil. Mag.,* 2007, **87**, 2561
30. D. Gryaznov, E. Heifets, E.A. Kotomin (in preparation).
31. C. E. McNeilly, *J. Nucl. Mater.,* 1964, **11 (1)**, 53
32. T. Gouder, A. Seibert, L. Havela, J. Rebizant, *Surf. Sci.,* 2007, **601**, L77
33. Matzke, H. *Science of Advanced LMFBR Fuels* (North Holland:Amsterdam, 1986).
34. J. M. Haschke, T. H. Allen, and L. A. Morales, *Science,* **287**, 285 (2000).